\documentstyle[12pt,a4wide]{article}

\def\be{\begin{equation}}
\def\ee{\end{equation}}
\def\bea{\begin{eqnarray}}
\def\eea{\end{eqnarray}}
\def\lb{\label}

\begin{document}

\author{Bernard Linet \thanks{E-mail: linet@lmpt.univ-tours.fr} \\
\small Laboratoire de Math\'ematiques et Physique Th\'eorique \\
\small CNRS/UMR 6083, Universit\'e Fran\c{c}ois Rabelais \\
\small Parc de Grandmont, 37200 TOURS, France}
\title{\bf Black holes in which the electrostatic 
or scalar equation is solvable in closed form}
\date{}
\maketitle

\begin{abstract}
We show that the method used in the Schwarzschild black hole for finding 
the elementary solution of the electrostatic equation in closed form
cannot extend in higher dimensions. By contrast, we prove the 
existence of static, spherically symmetric geometries with a 
non-degenerated horizon in which the static scalar equation
can be solved in closed form. We give the explicit results in 6 dimensions.
We determine moreover the expressions of the electrostatic potential and of the static scalar field 
for a point source in the extremal Reissner-Nordstr\"{o}m black holes in
higher dimensions. 
\end{abstract}

\thispagestyle{empty}

\section{Introduction}

In the Schwarzschild and Reissner-Nordstr\"{o}m black holes, 
it is astonished that the elementary solution in the Hadamard sense
of the electrostatic equation was found in closed form
\cite{copson1,copson2}. Hence in these background geometries the 
electrostatic potential generated by an electric point charge held fixed 
can be explicitly determined \cite{linet1,leaute}.
Likewise, the expression of the static scalar potential in these black holes is known
\cite{linet2}, see also \cite{mayo,wiseman}.
That seems miraculous since these equations are nonlinear partial differential
equations. In this work, we wonder from which this possibility comes and,
taking into account our analysis, whether it would be possible to
find the elementary solution
of the electrostatic or scalar equation in static, spherically symmetric geometries
in higher dimensions. 
More specifically we are interested in the black holes with a non-degenerated horizon.
We also examine these explicit determinations in the case of the extremal 
black holes in general relativity in higher dimensions.

A static, spherically symmetric geometry in $n+1$ dimensions can be 
described by a metric expressed in isotropic coordinates $(t,x^i)$,
$i=1,\ldots , n$. Setting 
$r=\sqrt{\left( x^1\right)^2+\cdots \left( x^n\right)^2}$, 
the electrostatic or scalar equation has the following form:
\be \lb{01}
\triangle A+h(r)\frac{x^i}{r}\partial_i A=0
\ee 
in which $h$ is calculated in terms of the components of the metric and
$\triangle$ is the Laplacian operator in $n$ dimensions.

We recall the definition of the elementary solution $A_e$ in the Hadamard sense
to equation (\ref{01}). Points $x$ and $x_0$ being given, we set
\be \lb{ga}
\Gamma (x,x_0)=( x^1-x_{0}^{1})^2+\cdots +( x^n-x_{0}^{n})^2 .
\ee
In the case where $n$ is odd, the elementary solution to equation (\ref{01}) 
is defined by the following development: 
\be \lb{03}
A_e(x,x_0)=\frac{1}{\Gamma^{n/2-1}}
\left( U_0(r,r_0)+U_1(r,r_0)\Gamma +\cdots \right) .
\ee
In the case where $n$ is even, it is defined by a development which contains
eventually logarithmic terms
\be \lb{03l}
A_e(x,x_0)=\frac{1}{\Gamma^{n/2-1}} \left( U_0(r,r_0)+U_1(r,r_0)\Gamma +\cdots 
\right) +\ln \Gamma \left( V_0(r,r_0)+V_1(r,r_0)\Gamma +\cdots \right) . 
\ee
The coefficients $U_n$ and $V_n$ are symmetric in $r$ and $r_0$. 
The first $U_0$ is obtained by the relation
\be \lb{u0}
2U_0'+hU_0=0
\ee
and the next coefficients by a recurrence relation. So, the elementary
solution $A_e$ is defined up to a multiplicative constant. 
 
The purpose of the present work is to determine the functions $h$ so that
the elementary solution to equation (\ref{01}) has the form
\be \lb{02}
A_e(x,x_0)=g(r)g(r_0)F\left( \frac{\Gamma (x,x_0)}{k(r)k(r_0)}\right)
\ee
where the functions $g$, $k$ and $F$ are to be determined.
We will show that a limited number of functions $k$ is possible. We will obtain
\be \lb{k}
k(r)=\frac{r^2-a^2}{b} , \quad k(r)=\frac{r^2+a^2}{b} , \quad k(r)=\frac{r^2}{b} , 
\quad k(r)=b ,
\ee
with $b\neq0$. In the first case, the range of variation of the coordinate $r$ 
is respectively $a\leq r <\infty$ or $0\leq r\leq a$ and in the three last cases
$r\geq 0$.

It will be convenient in the next to introduce the positive function $l$ defined by
\be \lb{0l}
h=\frac{l'}{l} ,
\ee
which is defined up to a multiplicative constant, and the function $\xi$ by setting
\be \lb{dxi}
\xi (r)=\frac{1}{g(r)} .
\ee

The paper is organized as follows. In section 2, we give the differential
equation verified by $h$ for each function $k$ enumerated in (\ref{k}). 
We also give the differential equation verified by $F$ and we obtained $g$
in terms of the function $l$ defined by (\ref{0l}). We recall in section 3 the geometric properties
of non-degenerated black holes in $n+1$ dimensions in the isotropic coordinates.
In section 4, we analyse the electrostatic equation in these black holes. 
Form (\ref{02}) of the elementary solution is only possible for $n=3$ and we find the known results for
the Schwarzschild and Reissner-Nordstr\"{o}m black holes. 
In section 5, we analyse the static scalar equation in these black holes. 
Form (\ref{02}) of the elementary solution is now possible in higher dimensions and we determine the functions
$l$, $F$ and $g$ when $n$ is odd. We illustrate our method with the case 
$n=5$. We furthermore solve in section 6 the electrostatic or scalar equation in the 
extremal Reissner-Nordstr\"{o}m black holes in $n+1$ dimensions. We add some 
concluding remarks in section 7.  

\section{System of differential equations}

By inserting form (\ref{02}) into equation (\ref{01}) and by using the relations 
$$
\partial_i\Gamma =2\left( x^i-x_{0}^{i}\right) ,\quad
x^i\partial_i\Gamma =\Gamma +r^2-r_{0}^{2} , \quad
\delta^{ij}\partial_i\Gamma \partial_j\Gamma =4\Gamma ,\quad
\delta^{ij}\partial_{ij}\Gamma =2n ,
$$
we obtain
\bea \lb{a1}
& & \left\{ g''+\frac{n-1}{r}g'+hg'\right\} F + \nonumber \\
& & \left\{ \frac{\Gamma}{kk_0} \left[ 
2g'\left( \frac{1}{r}-\frac{k'}{k}\right) 
+ g\left( -\frac{2k'}{kr}-\frac{k''}{k}+
\frac{2k'^2}{k^2}-\frac{(n-1)k'}{kr}\right)
+hg\left( \frac{1}{r}-\frac{k'}{k}\right) \right] \right. +\nonumber \\
& & \left. \quad 2g'\frac{r^2-r_{0}^{2}}{kk_0r} +g\left( \frac{2n}{kk_0}-
\frac{2(r^2-r_{0}^{2})k'}{k^2k_0r}\right) +hg\frac{r^2-r_{0}^{2}}{kk_0r}
\right\} F' + \nonumber \\
& & \left\{ \left( \frac{\Gamma}{kk_0}\right)^2 
\left[ g\left( \frac{k'^2}{k^2}
-\frac{2k'}{kr}\right) \right]  
+\frac{\Gamma}{kk_0} \left[ g\left( \frac{4}{kk_0}-
\frac{2k'(r^2-r_{0}^{2})}
{k^2k_0r} \right) \right] \right\} F''=0 
\eea
where $k_0$ denotes $k(r_0)$.
By considering the coefficient of $F''$ in (\ref{a1}), we see that a necessary condition
to derive a differential equation for $F$, with the variable 
$s$ defined by
\be 
s=\frac{\Gamma (x,x_0)}{k(r)k(r_0)} ,
\ee
is that the quantity $4/kk_0-2k'(r^2-r_{0}^{2})/k^2k_0r$ does not depend on 
$r_0$. It is not hard to show that this condition leads to expressions (\ref{k})
of the function $k$. We must now substitute these functions $k$ in (\ref{a1}) 
to verify whether it is possible to derive the differential equations verified by 
$h$, $g$ and $F$.

\subsection{Case I: $k(r)=(r^2-a^2)/b$}

With the choice $k(r)=(r^2-a^2)/b$, equation (\ref{a1}) becomes
\bea \lb{ab1}
& & \left\{ g''+\frac{n-1}{r}g'+hg' \right\} F + \nonumber \\
& & \left\{ s\left[ -2g'\frac{r^2+a^2}{r\left( r^2-a^2\right)}+
g\left( -\frac{2n+4}{r^2-a^2}+\frac{8r^2}{\left( r^2-a^2\right)^2}\right)
-hg\frac{r^2+a^2}{r\left( r^2-a^2\right)}\right] + \right. \nonumber \\
& & \left. \quad \frac{b^2}{r_{0}^{2}-a^2}
\left( \frac{2}{r}g'+\frac{2n-4}{r^2-a^2}g +\frac{1}{r}hg \right) 
+\frac{b^2}{r^2-a^2}\left( -\frac{2}{r}g'+\frac{4}{r^2-a^2}g
-\frac{1}{r}hg \right) \right\} F' + \nonumber \\
& & \frac{4gb^2}{\left( r^2-a^2\right)^2}\left\{ 
\frac{a^2}{b^2}s^2+s \right\} F'' =0 .
\eea
By seeing (\ref{ab1}), we can derive a differential  equation for
$F$, having the coefficients independent of $r_0$, by imposing
\be \lb{ab4}
g''+\frac{n-1}{r}g'+hg' =\frac{Ab^2}{\left( r^2-a^2\right)^2}g ,
\ee
\be \lb{ab5}
-2g'\frac{r^2+a^2}{r\left( r^2-a^2\right)}+g \left( -\frac{2n+4}{r^2-a^2}
+\frac{8r^2}{\left( r^2-a^2\right)^2}\right)
-hg\frac{r^2+a^2}{r\left( r^2-a^2\right)}=\frac{Cb^2}{\left( r^2-a^2\right)^2}g ,
\ee
\be \lb{ab6}
\frac{2}{r}g'+\frac{2n-4}{r^2-a^2}g+hg\frac{1}{r}=0 ,
\ee
\be \lb{ab7}
-\frac{2b^2}{r\left( r^2-a^2\right)}g'+\frac{4b^2}{\left( r^2-a^2\right)^2}g
-\frac{b^2}{r\left( r^2-a^2\right)}hg=\frac{Bb^2}{\left( r^2-a^2\right)^2}g ,
\ee
where the constants $A$, $B$ and $C$ are arbitrary.
Equation (\ref{ab1}) becomes thereby
\be \lb{ab8}
g\frac{b^2}{\left( r^2-a^2\right)^2}\left[ 4\left( \frac{a^2}{b^2}s^2
+s\right) F''+(Cs+B)F'+AF\right] =0 .
\ee

The system of differential equations (\ref{ab4}-\ref{ab7}) is overdetermined but by taking the following values for 
the constants: $B=2n$ and $C=4n a^2/b^2$,
the two equations (\ref{ab5}) and (\ref{ab7}) can be deduced from
(\ref{ab4}) et (\ref{ab6}). Now, the differential equation (\ref{ab8}) for $F$ is 
\be \lb{ab9}
4\left( \frac{a^2}{b^2}s^2+s\right) F''+\left( 4n\frac{a^2}{b^2}s+2n \right) F'
+AF =0 .
\ee
We must choose a solution to (\ref{ab9}) which yields a development of $A_e$ 
in the Hadamard sense (\ref{03}) or (\ref{03l}).

Now, we derive (\ref{ab6}) with respect to $r$ then  
by combining with (\ref{ab4}), we obtain
the decoupled differential equation verified by $h$
\be \lb{ab10}
h'+\frac{1}{2}h^2+\frac{n-1}{r}h-2\left( n(n-2)a^2-Ab^2\right)
\frac{1}{\left( r^2-a^2\right)^2} =0 .
\ee
When $h$ is known, we can solve equation (\ref{ab6}) in terms of the
fonction $l$ defined by (\ref{0l}) to determine $g$ as a positive function
\be \lb{ab6l}
g(r)=\pm \frac{1}{\left( r^2-a^2\right)^{n/2-1}\sqrt{l(r)}}.
\ee
Finally, the system of differential equations (\ref{ab4}-\ref{ab7}) is verified.

It is more convenient to study the Riccati equation (\ref{ab10}) by using
an equivalent linear differential equation of the second order. The usual manner
is to consider the function $w$, defined by  $h=2w'/w$, which obeys
$$
w''+\frac{n-1}{r}w'-\frac{n(n-2)a^2-Ab^2}{(r^2-a^2)^2}w =0 .
$$
However, we will prefer to use directly the function $\xi$, inverse of $g$, related to $h$ 
through (\ref{ab6l}). We find 
\be \lb{32}
\xi ''+\left( \frac{n-1}{r}-\frac{2(n-2)r}{r^2-a^2}\right) \xi '
+\frac{Ab^2}{(r^2-a^2)^2}\xi =0 .
\ee

Without lost of generality we may take $b=a$ since this corresponds to a
simple change of the function $F$ and of the constant $A$.

\subsection{Case II: $k(r)=(r^2+a^2)/b$}

In this case, we obtain the formulas by performing the substitution $-a^2$ to $a^2$ in 
the expressions (\ref{ab9}-\ref{32}). We can also take $b=a$.

\subsection{Case III: $k(r)=r^2/b$}

We obtain the formulas by putting $a=0$ in the expressions given in the case I. Without lost
of generality, we take $b=1$ in order to carry out the calculations. 
From (\ref{ab9}) we have for $F$
\be \lb{f3}
4sF''+2nF'+AF=0 .
\ee
The differential equation (\ref{ab10}) verified by $h$ takes the form 
\be 
h'+\frac{1}{2}h^2+\frac{n-1}{r}h+\frac{2A}{r^4}=0 .
\ee
The function $g$ is determined by the relation
\be \lb{iiil}
g(r) =\frac{1}{r^{n-2}\sqrt{l(r)}} .
\ee
Finally, equation (\ref{32}) reduces to
\be \lb{323}
\xi ''+\frac{3-n}{r}\xi '+\frac{A}{r^4}\xi =0 .
\ee

\subsection{Case IV: $k(r)=b$}

Without lost of generality we take $b=1$ in order to carry out the calculations.
We return to equation (\ref{a1}) which takes the form
\bea \lb{ak1}
& & \left\{ g''+\frac{n-1}{r}g' +hg' \right\} F+ \nonumber \\
& & \left\{ s\left[ \frac{2}{r}g' +\frac{1}{r}hg \right] + 
2g'\frac{r^2-r_{0}^{2}}{r}+2ng+hg\frac{r^2-r_{0}^{2}}{r}
\right\} F' + \nonumber \\
& & 4gsF'' =0 .
\eea
The coefficient of $F'$ in (\ref{ak1}) does not depend on $r_0$ if we put 
\be \lb{ak2}
2g'+hg=0 .
\ee
Then, we obtain a differential equation for $F$ by putting
\be \lb{ak3}
g''+\frac{n-1}{r}g'+hg'=Ag
\ee
where the constant $A$ is arbitrary.
Consequently, equation (\ref{ak1}) reduces to
\be \lb{ak4}
4sF''+2nF'+AF=0 .
\ee
which coincides with (\ref{f3}).

From  (\ref{ak2}), we obtain the relation 
\be \lb{ak2l}
g(r) =\frac{1}{\sqrt{l(r)}} .
\ee
By combining (\ref{ak2}) and (\ref{ak3}) we derives the differential equation
verified by $h$
\be \lb{ak5}
h'+\frac{1}{2}h^2+\frac{n-1}{r}h+2A=0 .
\ee

We can also study the Riccati equation (\ref{ak5}) by a
linear differential equation for $\xi$
related to $h$ by (\ref{ak2l}) which is
\be \lb{ak6}
\xi ''+\frac{n-1}{r}\xi '+A\xi =0 .
\ee

\section{Geometric properties of non-degenerate black holes}

The metric describing a static, spherically symmetric black hole with
the topology $I\! \! R^2\times S^{n-1}$ is expressed in standard coordinates
$(t,R,\theta_1,\ldots ,\theta_{n-2},\varphi )$ under the form
\be  \lb{m1}
ds^2=g_{tt}(R)dt^2+g_{RR}(R)dR^2+R^2d\Omega_{n-1}^{2} ,
\ee
where $d\Omega_{n-1}^{2}$ represents the metric of the unit sphere $S^{n-1}$. A 
non-degenerate horizon at $R=R_H$, $R_H>0$, is located at a simple zero of $g_{tt}$ 
and at a simple pole of $g_{RR}$. In these coordinates the metric is well defined for
$R>R_H$. Hence we have the behaviors of the components at $R=R_H$
\be \lb{m2}
g_{tt}(R)\sim -g_{0}^{2}\left( R-R_H\right) ,\quad
g_{RR}(R)\sim \frac{g_{1}^{2}}{R-R_H} \quad {\rm as}\; R\rightarrow R_H .
\ee

We are going to write down metric (\ref{m1}) in isotropic coordinates. 
We define the radial coordinate $r$ by the differential relation
\be \lb{m4}
\sqrt{g_{RR}(R)}\frac{dR}{dr}=\frac{R}{r} ,
\ee
where $r$ is defined up to a multiplicative constant. We choose the solution to (\ref{m4}) 
so that $R\sim r$ as $R\rightarrow \infty$. Metric (\ref{m1}) takes the form
\be \lb{m6}
ds^2=-N^2(r)dt^2+B^2(r)\left( dr^2+r^2d\Omega_{n-1}^{2}\right) ,
\ee
with
\be \lb{m5}
N(r)=\sqrt{-g_{tt}(R(r))} , \quad B(r)=\frac{R(r)}{r} .
\ee

Since a non-degenerate black hole has behaviors (\ref{m2}), 
formula (\ref{m4}) can be written
\be \lb{m4a}
\frac{g_1dR}{R_H\sqrt{R-R_H}} \left( 1+O(R-R_H)\right) =\frac{dr}{r} .
\ee 
According to (\ref{m4a}), the value $R=R_H$ is reached for a finite value of $r$ 
strictly positive, denoted $r_H$. By performing the integration, we get
\be \lb{m8}
\frac{2g_1}{R_H}\sqrt{R-R_H}\sim \frac{1}{r_H}\left( r-r_H\right) \quad {\rm as}\; 
r\rightarrow r_H .
\ee
Substituting (\ref{m8}) into (\ref{m5}), we obtain the following behaviors:
\be \lb{m9}
N(r)\sim \frac{g_0R_H}{2g_1r_H} 
\left( r-r_H\right) , \quad B(r)\sim \frac{R_H}{r_H}
\quad {\rm as}\; r\rightarrow r_H .
\ee
The metric will be asymptotically flat
if $N(r)\sim 1$ and $B(r)\sim 1$ as $r\rightarrow \infty$.

We turn to the converse. A geometry described by metric (\ref{m6}) has a
non-degenerate horizon at $r=r_H$, $r_H>0$, if we have the behavior at $r=r_H$
\be \lb{m9n}
N(r)=\kappa B(r_H)\left( r-r_H\right) \left( 1+O(r-r_H)\right) , 
\ee
where $\kappa$ is a nonvanishing constant, called the surface gravity,  
and the behavior at $r=r_H$
\be \lb{m9b}
B(r)=B(r_H)-\frac{B(r_H)}{r_H}\left( r-r_H\right) +
\frac{B''(r_H)}{2}\left( r-r_H\right)^2+O\left( (r-r_H)^3\right) ,
\ee
with the assumptions $B(r_H)\neq 0$ and $B''(r_H)r_H+2B(r_H)\neq 0$.
The proof is derived by means of the transformation of radial coordinates
$R(r)=B(r)r$ which has to have from (\ref{m8}) the behavior 
$R(r)-R_H\sim {\rm const.}\; (r-r_H)^2$ as $r\rightarrow r_H$ where the 
constant does not vanish.

In the Cartesian coordinates $(x^i)$ associated to 
$(r,\theta_1,\ldots ,\theta_{n-2},\varphi )$, metric (\ref{m6}) becomes
\be \lb{m7}
ds^2=-N^2(r)dt^2+B^2(r)
\left( \left(dx^1\right)^2+\cdots +\left( dx^{n-1}\right)^2\right) .
\ee
The electrostatic or scalar equation (\ref{01}) has been written 
down in the background metric (\ref{m7}).

\section{Solvable electrostatic equation}

In the background metric (\ref{m7}), the equation for the time component
of the electromagnetic potential, denoted $V$ in the static case, can be deduced from the
Maxwell equations. We obtain equation (\ref{01}) where $h_{em}$ is
calculated by (\ref{0l}) from the positive function
\be \lb{e1}
l_{em}(r)=\frac{B^{n-2}(r)}{N(r)} .
\ee
For an electric point charge $q$ at $x=x_0$, the source of equation (\ref{01})
is
\be \lb{sem}
-s_{n-1}\frac{q}{l_{em}(r_0)}\delta^{(n)}(x,x_0)
\ee
where  $\delta^{(n)}$ is the Dirac distribution in $I\! \! R^n$
and where $s_{n-1}$ denotes the area of the unit sphere $S^{n-1}$. 
Taking into account (\ref{u0}), we consider the elementary solution $V_e$
which has a first coefficient $U_0$ given by
\be \lb{u0n}
U_0(r,r_0)=\frac{q}{\sqrt{l_{em}(r)}\sqrt{l_{em}(r_0)}} .
\ee

For a non-degenerated black hole, we have behaviors (\ref{m9})
and we thus obtain from (\ref{e1}) the behavior at $r=r_H$
\be \lb{e2}
l_{em}(r)=\frac{l_0}{r-r_H} \left( 1+O(r-r_H)\right) .
\ee
We can therefore hope to find a solution solvable in form (\ref{02})
by considering the case I of section 2 with $a=r_H$ with the range of coordinate
$r\geq a$ in which it should be found $l_{em}(r)$ positive for $r>a$.   

Since we intend to discuss equation (\ref{32}) for $\xi$ defined by (\ref{ab6l}), 
we evaluate from (\ref{e2}) its behavior at $r=r_H$ 
\be \lb{e3}
\xi_{em}(r)= \xi_0 \left( r-r_H\right)^{(n-3)/2} \left( 1+O(r-r_H)\right) .
\ee
The problem is to know if the differential equation (\ref{32})
verified by $\xi$ admits solutions with behavior (\ref{e3}) at $r=r_H$.
Moreover $\xi_{em}$ must be a positive function for $r>r_H$.

\subsection{Analysis of the behavior at $r=a$}

We start by studing the behavior of the solutions to equation (\ref{32}) at $r=a$. 
To do this, we introduce the variable $y$ by
\be \lb{33}
y=\frac{r}{a} .
\ee
Equation (\ref{32}) takes the standard form to analyse the behavior of the
solutions at $y=1$
\be \lb{as1}
\xi ''+p(y)\xi '+q(y)\xi =0
\ee
in which $p$ et $q$ are series in power of $(y-1)$ given by
\be \lb{a2}
p(y)=\sum_{p=0}^{\infty}p_p(y-1)^{p-1} ,\quad
q(y)=\sum_{p=0}^{\infty}q_p(y-1)^{p-2} ,
\ee
with
$$
p_0=2-n, \; p_1=\frac{n}{2}, \; p_2=\frac{2-3n}{4}, \; \ldots ,\quad
q_0=\frac{A}{4}, \; q_1=-\frac{A}{4}, \; q_2=\frac{3A}{16}, \; \ldots
$$
According to (\ref{a2}) the point $y=1$ is a singular regular point of equation (\ref{as1}). 
The indicial equation relative to $y=1$ is
\be \lb{34}
c^2+(1-n)c+\frac{A}{4}=0 .
\ee
We denote $c_1$ and $c_2$ the roots of the indicial equation (\ref{34}) and it is useful
to introduce the difference $s=c_1-c_2$.

In the present case, we must confine ourselves to real roots $c_1$ and $c_2$
and consequently we must assume that $A\leq (n-1)^2$. We indicate $c_1$ largest
of the two roots and we get $c_2=n-1-c_1$ with  $c_1\geq (n-1)/2$.
The difference $s$ is then given by $s=2c_1+1-n$ with $s\geq 0$.

There is always a solution $\xi_1$ which has the exponent $c_1$ at $y=1$ and which is expressed
in the form of the series 
\be \lb{reg}
\xi_1(y)=\sum_{p=0}^{\infty} a_p(y-1)^{c_1+p} .
\ee
We can in general seek the solution $\xi_2$ which has the exponent
$c_2$ at $y=1$ in the form of the series
\be \lb{38}
\xi_2(y)=\sum_{p=0}^{\infty}b_{p}(y-1)^{c_2+p} .
\ee
The recurrence relation for the coefficients $b_p$ of the series (\ref{38}) is
\be \lb{38r}
p(p-s)b_{p}=-\sum_{m=0}^{p-1}b_{m}\left[ (m+c_2)p_{p-m}+q_{p-m}\right] ,\quad p\geq 1 .
\ee

However, when $s$ is equal to an integer, we see an obstruction for determining $b_s$ 
by the recurrence relation (\ref{38r}) when the right side of this relation 
does not vanish.
If it is not zero then the second solution $\xi_2$ is not given by the series (\ref{38})
but by the following expansion
\be \lb{39}
\xi_2(y)=\xi_1(y)\ln (y-1)+\sum_{p=0}^{\infty}d_p(y-1)^{c_2+p} 
\ee
which contains logarithmic terms. Of course, solution (\ref{39}) has not the required behavior
at $y=1$. 

In the electrostatic case, the solution $\xi$ must have behavior (\ref{e3}) at
$r=a$. Since $c_1\geq (n-1)/2$, this is $\xi_2$ which must have
this behavior. We thus put
\be \lb{59}
c_1=\frac{n+1}{2} , \quad c_2=\frac{n-3}{2} , 
\ee
and $A=(n+1)(n-3)$. With choice (\ref{59}), we get $s=2$.
The recurrence relation (\ref{38r}) for $p=2$ is
$$
0=b_0[c_2p_2+q_2]+b_1[(1+c_2)p_1+q_1] \quad {\rm with}\; b_1=b_0[c_2p_1+q_1] .
$$
By inserting the $p_k$ and $q_k$ determined in (\ref{a2}) for choice (\ref{59}), we find
\be \lb{510}
0=b_0\frac{4n-3-n^2}{4} .
\ee
Equation (\ref{510}) shows that the solution $\xi_2$ has the form (\ref{38}) if and only if
$4n-3-n^2=0$. 

In consequence, the solution $\xi_2$ cannot have behavior (\ref{e3}) except
in the particular case $n=3$ for which $c_1=2$ and $c_2=0$. In this case $A=0$,
the solutions  (\ref{511g}) given in appendix A reduces to
\be \lb{511s}
\xi_1(r)=\frac{(r-a)^2}{r}, \quad \xi_2(r)=\frac{(r+a)^2}{r} .
\ee
However, we satisfy behavior $\xi (r)\sim \xi_0$ as $r\rightarrow a$ by taking the solution
\be \lb{511}
\xi_{em}(r)=\tau_1\xi_1(r)+\tau_2\xi_2(r)
\ee
where the constants $\tau_1$ and $\tau_2$ ensure that $\xi_{em}$ is a
positive function for $r>a$.

\subsection{Reissner-Norstr\"{o}m black holes in four dimensions}

We are going to check that we can treat the electrostatic equation in the Reissner-Norstr\"{o}m black holes
with solution (\ref{511}). It is parametrized by the mass $m$ and the electric charge $e$
satisfying $\vert e\vert < m$.  In standard coordinates, its metric is 
\be \lb{rn1}
ds^2=-\left( 1-\frac{2m}{R}+\frac{e^2}{R^2}\right) dt^2+
\left( 1-\frac{2m}{R}+\frac{e^2}{R^2}\right)^{-1}dR^2+R^2d\Omega_{2}^{2} ,
\ee
with $R>R_H$ where $R_H$ is the exterior horizon $R_H=m+\sqrt{m^2-e^2}$.
The change (\ref{m4}) of the radial coordinates gives
\be \lb{rn2}
R(r)=r+m+r_{H}^{2}\frac{1}{r}
\ee
where $r_H=\sqrt{m^2-e^2}/2$ satisfying the condition $m>2r_H$.
The metric expressed in isotropic coordinates is thus characterized by 
\be \lb{rn3}
N(r)=\frac{r^2-r_{H}^{2}}{r^2+mr+r_{H}^{2}} ,\quad
B(r)=\frac{r^2+mr+r_{H}^{2}}{r^2} .
\ee
From (\ref{e1}) we deduce
\be \lb{rn4}
l_{em}(r)=\frac{\left( r^2+mr+r_{H}^{2}\right)^2}{r^2(r+r_H)(r-r_H)} .
\ee
Consequently from (\ref{ab6l}), we get
\be \lb{rn5}
\xi_{em}(r)=\frac{r^2+mr+r_{H}^{2}}{r} .
\ee
Expression (\ref{rn5}) coincides with (\ref{511}) for a judicious choice of
$\tau_1$ and $\tau_2$.
In the case $A=0$, expression (\ref{27}) reduces to
\be \lb{rn8}
F_{em}(s)=\frac{1+2s}{\sqrt{s(s+1)}} .
\ee
The elementary solution which is normalized by (\ref{u0n}) has the expression
\be \lb{rne}
V_e(x,x_0)=qg_{em}(r)g_{em}(r_0)r_HF_{em}
\left(\frac{r_{H}^{2}\Gamma (x,x_0)}{(r^2-r_{H}^{2})(r_{0}^{2}-r_{H}^{2})}\right)  ,
\ee
where $g_{em}=1/\xi_{em}$.

The determination of the electrostatic potential generated by a point charge (\ref{sem})
from the elementary solution (\ref{rne}) is well known \cite{linet1, leaute}. 
Without details, we give the result
\be \lb{rn9}
V(x)=V_e(x,x_0)+qmg_{em}(r)g_{em}(r_0) 
\ee
where $g_{em}$ is indeed the electrostatic monopole.
The electrostatic potential (\ref{rn9}) gives an electric field which is regular at
the horizon $r=r_H$ and at the infinity.

\section{Solvable scalar equation}

In the background metric (\ref{m7}), the static scalar equation for a scalar field $\psi$ 
minimally coupled is written with a function $h_{sc}$ calculated by (\ref{0l}) from the positive function
\be \lb{5a}
l_{sc}(r)=N(r)B^{n-2}(r) .
\ee
For a scalar charge $g$ located at $x=x_0$, the source of equation (\ref{01}) is
\be \lb{ssc}
-s_{n-1}\frac{N(r_0)}{l_{sc}(r_0)}\delta^{(n)}(x,x_0) .
\ee
We consider the elementary solution $\psi_e$ which has the first coefficient
\be \lb{s0n}
U_0(r,r_0)=\frac{g}{\sqrt{l_{sc}(r)}\sqrt{l_{sc}(r_0)}} .
\ee

For a non-degenerated black hole, behavior (\ref{m9}) at $r=r_H$ yields
\be \lb{5b}
l_{sc}(r)=l_0 (r-r_H)\left( 1+O(r-r_H)\right) .
\ee
From (\ref{5b}) we deduce the equivalent behavior for $\xi$ at $r=r_H$
\be \lb{5c}
\xi_{sc}(r)=\xi_0 (r-r_H)^{n/2-1/2}\left( 1+O(r-r_H)\right) .
\ee

\subsection{Analysis of the behavior at $r=a$}

We proceed as in the previous section to analyse
the behavior of the solutions to equation (\ref{01})
at $r=a$. In the present case, the solution $\xi$ must have the behavior (\ref{5c}).
We can always take $\xi_1$ with such a behavior by putting
\be \lb{5d}
c_1=\frac{n-1}{2} ,\quad c_2=\frac{n-1}{2} , 
\ee
and $A=(n-1)^2$. 
We have $s=0$ and the second solution $\xi_2$
will contain logarithmic terms.

By introducing a function $f$ by
\be \lb{5f}
\xi (r)=\frac{(r^2-a^2)^{(n-1)/2}}{r^{(n-2)}}f(r) ,
\ee
and a new variable $x$ by
\be \lb{5g}
x=\frac{r^2}{a^2} ,
\ee
the differential equation (\ref{32}) takes the form
\be \lb{5h}
x(x-1)f''+\left( (3-n/2)x-(2-n/2)\right) f'+(3/4-n/4)f =0 .
\ee
Equation (\ref{5h}) is the hypergeometric equation with the coefficients
$\alpha =1/2$, $\beta = -n/2+3/2$ and $\gamma =2-n/2$.
A solution $f_1$ to (\ref{5h}) which is regular at $x=1$ corresponds to
the solution $\xi_1$ having the exponent $(n-1)/2$ at  $r=a$.

\subsection{Scalar potential in $n+1$ dimensions}

There is a marked contrast with respect to the electrostatic case. There exists geometries
having non-degenerated horizon in which the elementary solution of the static scalar equation
has form (\ref{02}). Without determining them explicitly, we can prove certain general properties. 

In the case where $n$ is even, $n\geq 4$, the coefficient $\gamma$ 
of the hypergeometric equation (\ref{5h}) is a negative integer or zero and the solutions
do not seem expressible in terms of the elementary functions. In the case
where $n$ is odd, $n\geq 3$, the coefficient $\beta$ is equal to a negative integer or zero.
We set $n=2p+1$. A solution to (\ref{5h}) is the hypergeometric function defined by
a series of power of $x$. However, this series reduces for $\beta =1-p$ to a polynome
of degre $p-1$. Its expression is given by
\be \lb{hgp}
f_1(x)=1+\sum_{r=1}^{p-1}\frac{(1-p)_r(1/2)_r}{(3/2-p)_r}\frac{x^r}{r!}
\ee  
where $(\; )_r$ denotes the Pochhammer symbol $(a)_r=a(a+1)\cdots (a+r-1)$.
Solution (\ref{hgp}) is obviously regular at $x=1$ and leads to the desired solution $\xi_1$. 
However, we must prove that $f_1(x)$ is strictly positive for $x\geq 1$. This results from
the fact that 
each ratio $(1-p)_r/(3/2-p)_r$ is positive for $p\geq r+1$ and so all coefficients of
polynome (\ref{hgp}) are positive.

We must also determine the function $F$ that we write here $F_n$ in $n+1$ dimensions.
Since $A=(n-1)^2$, equation (\ref{ab9}) is read
\be \lb{a}
4s(s+1)F_n''+(4ns+2n)F_n'+(n-1)^2F_n=0 .
\ee
By differentiating equation (\ref{a}), we obtain the recurrence relation
\be \lb{b}
F_{n+2}(s)=-\frac{2}{n-2}\frac{d}{ds}F_n(s) .
\ee
Obviously, differentiation (\ref{b}) preserves the development of $F$ in the Hadamard sense.

Setting $\tau =-s$ in (\ref{a}), we obtain
\be \lb{c}
\tau (\tau -1)F_n''+(n\tau -n/2)F_n'+(n-1)^2/4\; F_n=0 
\ee
which is the hypergeometric equation with coefficients
$\alpha =(n-1)/2$, $\beta =(n-1)/2$ and $\gamma =n/2$.
As in the previous case for $f$, the solutions $F_n$ to (\ref{c})
are expressible in closed form only for $n$ odd. We confine henceforth in this case. 

Taking into account (\ref{b}), it is sufficient to know $F_3$ having a development in the Hadamard sense. 
Its expression is given by (\ref{27}) with $A=4$ 
\be \lb{5k}
F_3(s)=\frac{1}{\sqrt{s(s+1)}} 
\ee
which has the typical development (\ref{exp}) with $A=4$.
By applying successively (\ref{b}) on $F_3$, we see that $F_{2p+1}$ has the expression
\be \lb{5kp}
F_{2p+1}(s)=\frac{(-2)^{p-1}}{1.3\ldots (2p-3)}\frac{d^{p-1}}{ds^{p-1}}F_3(s) .
\ee
According to (\ref{5k}), we find the two following behaviors for $F_{2p+1}$
\be \lb{simh}
F_{2p+1}(s)\sim \frac{1}{s^{p-1/2}} \quad {\rm as} \; s\rightarrow 0 ,
\ee
\be \lb{simi}
F_{2p+1}(s)\sim \frac{2^{p-1}(p-1)!}{1.3\ldots (2p-3)s^p} \quad {\rm as} \; s\rightarrow \infty .
\ee

We now give the expression of the elementary solution $\psi_e$ 
in the case $n=2p+1$ 
\be \lb{e}
\psi_e(x,x_0)=\frac{g}{\xi_1(r)\xi_1(r_0)}F_{2p+1}(s)
\ee
with
$$
s=\frac{a^2\Gamma (x,x_0)}{(r^2-a^2)(r_{0}^{2}-a^2)} .
$$
Let us prove that the elementary solution (\ref{e}) is regular at the horizon $r=a$.
From (\ref{simi}) we have
$$
F_{2p+1}(s) \sim {\rm const.}\; \frac{(r-a)^p}{\left[ \Gamma(x_H,x_0)\right]^p} \quad {\rm as}\; r\rightarrow a ,
$$
where $x_H$ is a point $x$ with $r=a$. Now from (\ref{5f}) we have 
$$
\xi_1(r) \sim {\rm const.}\; (r-a)^p \quad {\rm as}\; r\rightarrow a ,
$$
therefore $\psi_e(x_H,x_0)$ is finite.  

We now seek the asymptotic behavior of $\psi_e$ for $r\rightarrow \infty$. The variable $s$ takes the finite value
$a^2/(r_{0}^{2}-a^2)$. Thus, $\psi_e$ tends to a spherically symmetric
field having the same asymptotic behavior than $1/\xi_1(r)$ as $r\rightarrow \infty$.

We point out the monopolar solution of the scalar homogeneous equation, denoted $\psi_0$, satisfies
$$
\frac{d}{dr}\psi_0(r) =\frac{1}{l_{sc}(r)r^{n-1}} .
$$
Taking into account behavior (\ref{5b}) for $l_{sc}$ at $r=a$, we see that $\psi_0$ 
tends to blow up as $r\rightarrow a$. Therefore, a possible addition of $\psi_0$ to 
the elementary solution (\ref{e}) does not give a regular solution at the horizon $r=a$.
In consequence the scalar potential $\psi$ generated by the scalar charge (\ref{ssc}) has the expression
\be \lb{esp}
\psi (x)=N(r_0)\psi_e(x,x_0) .
\ee

When $r_0\rightarrow a$ we obtain likewise that $\psi_e$ has a finite value.
Then, expression (\ref{esp}) of $\psi$ tends to zero due to the factor $N(r_0)$;
the scalar field $\psi$ disapears. We point out that the horizon $r=a$ is not
a surface of constant potential. 

\subsection{Examples for black holes in 4 and 6 dimensions}

In the case $n=3$, the solution $f_1$ given by (\ref{hgp}) reduces to a constant. From (\ref{5f})
we obtain  
\be \lb{5i}
\xi_1(r)=\frac{r^2-a^2}{r} ,
\ee
already given in (\ref{5a0}). The other solution $\xi_2$ contains a logarithmic term.
By substituting (\ref{5i}) in (\ref{ab6l}) we obtain 
\be \lb{5j}
l_{sc}(r)=\frac{r^2-a^2}{r^2} .
\ee

We verify that the calculation of $l_{sc}$ given by (\ref{5a}) for the Reissner-Nordstr\"{o}m metric (\ref{rn3}) 
gives (\ref{5j}). The function $F_{sc}$ is $F_3$ given by (\ref{5k}).
So, the elementary solution is
\be \lb{rnse}
\psi_e(x,x_0)=\frac{grr_0r_H}{(r^2-r_{H}^{2})(r_{0}^{2}-r_{H}^{2})}F_{sc}
\left( \frac{r_{H}^{2}\Gamma (x,x_0)}{(r^2-r_{H}^{2})(r_{0}^{2}-r_{H}^{2})}\right) .
\ee
It is regular at the horizon $r=r_H$ and at the infinity. 
The scalar field for the scalar charged source (\ref{ssc}) is
\be \lb{rns}
\psi (x)=\frac{r_{0}^{2}-r_{H}^2}{r_{0}^{2}+mr_0+r_{H}^{2}}\psi_e(x,x_0) .
\ee
Result (\ref{rns}) is already known \cite{linet2}.

In the case $n=5$, the solution $f_1$ given by (\ref{hgp}) reduces to $1+x$. From (\ref{5f})
we obtain 
\be \lb{5l}
\xi_1(r)=\frac{(r^2+a^2)(r^2-a^2)^2}{r^3} .
\ee
The other solution $\xi_2$ contains a logarithmic term
$$
\xi_2(r)=\frac{2a(r^2-a^2)^2}{r^2}+ \frac{(r^2+a^2)(r^2-a^2)^2}{r^3} \ln \left( \frac{r-a}{r+a}\right) .
$$
By considering only $\xi_1$ given by (\ref{5l}), we obtain from (\ref{ab6l})
\be \lb{5m}
l_{sc}(r)=\frac{(r^2-a^2)(r^2+a^2)^2}{r^6} .
\ee

The function $F_{sc}$ is calculated by using (\ref{5kp})
\be
F_{sc}(s)=\frac{2s+1}{(s(s+1))^{3/2}} .
\ee
We check that the elementary solution normalized by (\ref{s0n}) is
\be 
\psi_e(x,x_0)=\frac{ga^3}{\xi_1(r)\xi_1(r_0)}F_{sc}\left( \frac{a^2\Gamma (x,x_0)}{(r^2-a^2)(r_{0}^{2}-a^2)}\right) ,
\ee
which is regular at the horizon $r=a$ and at the infinity.
The scalar field generated by the scalar charged source (\ref{ssc}) is 
\be 
\psi (x)=N(r_0)\psi_e(x,x_0) .
\ee

We can wonder whether a simple metric describing a black hole gives expression
(\ref{5m}) for $l_{sc}$. According to (\ref{5a}) for $n=5$, we can for example take
the following components of the metric in isotropic coordinates
\be \lb{5n}
N(r)=\frac{r^2-a^2}{r^2+a^2} ,\quad B(r)=\frac{r^2+a^2}{r^2} .
\ee
By changing the radial coordinates
$$
R=\frac{r^2+a^2}{r} ,\quad r=\frac{R+\sqrt{R^2-4a^2}}{2} ,
$$
the corresponding metric (\ref{5n}) becomes
\be \lb{5o}
ds^2=-\left( 1-\frac{4a^2}{R^2}\right) dt^2+\left( 1-\frac{4a^2}{R^2}\right)^{-1}dR^2+R^2d\Omega_{4}^{2} ,
\ee
with $R>R_H$ where $R_H$ is the non-degenerated horizon located at $R_H=2a$. 

\section{Extremal black hole in $n+1$ dimensions}

We are going to verify that the case III with $A=0$ enables us to derive
the elementary solutions to equation (\ref{01}) 
in the extremal black holes of the Einstein-Maxwell theory at $n+1$ 
dimensions \cite{myers}. In standard coordinates the metric is written as
\be \lb{rn10}
ds^2=-\left( 1-\frac{R_{H}^{n-2}}{R^{n-2}}\right)^2 dt^2+
\left( 1-\frac{R_{H}^{n-2}}{R^{n-2}}\right)^{-2}dR^2+R^2d\Omega_{n-1}^{2} ,
\ee
with $R>R_H$ where $R_H$ is the degenerated horizon, $R_H>0$.
Transformation (\ref{m4}) of the radial coordinates gives
\be \lb{rn11}
R(r)=\left( r^{n-2}+R_{H}^{n-2}\right)^{1/(n-2)} .
\ee
This leads to characterize metric (\ref{m6}) by
\be \lb{rn12}
N(r)=\frac{r^{n-2}}{r^{n-2}+R_{H}^{n-2}} ,\quad
B(r)=\frac{\left( r^{n-2}+R_{H}^{n-2}\right)^{1/(n-2)}}{r} ,
\ee
where the horizon is located at $r=0$.

From (\ref{rn12}) we can calculate $l_{em}$  
\be \lb{rn13}
l_{em}(r)= \frac{\left( r^{n-2}+r_{H}^{n-2}\right)^2}{r^{2n-4}}
\ee
and according to (\ref{iiil}) we must have
\be \lb{rn14}
\xi_{em}(r)=r^{n-2}+R_{H}^{n-2} .
\ee

We are in a solvable case since (\ref{rn14}) is a solution to
equation (\ref{323}) with $A=0$. 
The function $F_{em}$ is the solution to equation (\ref{f3}) with $A=0$ ; we get 
\be \lb{rn16}
F_{em}(s)=\frac{1}{s^{n/2-1}} .
\ee

The elementary solution $V_e$ of the electrostatic equation in background metric (\ref{rn10}) is  
\be \lb{rn17}
V_e(x,x_0)=\frac{q}{\xi_1(r)\xi_1(r_0)}F_{em}
\left( \frac{\Gamma (x,x_0)}{r^2r_{0}^{2}}\right) .
\ee
We must discuss the regularity of the electrostatic field at the horizon located at $r=0$.
To do this, we calculate the electromagnetic invariant 
$$
F_{\mu \nu}F^{\mu \nu}=\frac{2}{N^2B^2}\delta^{ij}\partial_iV_e\partial_jV_e
$$
in the limit where $r$ tends to zero. Since $s\sim 1/r^2$ as $r\rightarrow 0$, we see 
that the electromagnetic invariant is bounded on the horizon. 
We turn now to the regularity at the infinity. From the Gauss theorem, the electric
flux through a sphere containing the charge $q$ is $s_{n-1}q$. But $V_e$
has the asymptotic behavior
$$
V_e(x) \sim \frac{qr_{0}^{n-2}}{r^{n-2}\left( r_{0}^{n-2}+R_{H}^{n-2} \right)} \quad 
{\rm as}\; r \rightarrow \infty 
$$
instead of $q/r^{n-2}$ as determined by the Gauss theorem.
Fortunately, it is possible to add to $V_e$ a monopolar solution of the homogeneous
electrostatic equation
which induces a regular electric field at the horizon and at the infinity. It is easy to
see that it is given by $1/\xi_1$.
In order to satisfy the Gauss theorem, the electrostatic potential generated by an electric 
charge $q$ is thereby
\be \lb{rn17b}
V(x)=\frac{q}{\left( 1+\frac{R_{H}^{n-2}}{r^{n-2}}\right) \left( 1+\frac{R_{H}^{n-2}}{r_{0}^{n-2}}\right) 
\Gamma^{n/2-1}(x,x_0)}
+\frac{qR_{H}^{n-2}}{r^{n-2}r_{0}^{n-2}\left( 1+\frac{R_{H}^{n-2}}{r^{n-2}}\right) 
\left( 1+\frac{R_{H}^{n-2}}{r_{0}^{n-2}}\right) } .
\ee
This generalizes in $n+1$ dimensions the known result for $n=3$ \cite{leaute, hanni}.

We can also treat the static scalar equation. From (\ref{5a}) we see that
$l_{sc}(r)=1$.
So,  $h_{sc}(r)=0$ and equation (\ref{01}) reduces to the Laplacian in $n$ dimensions.
According to source (\ref{ssc}), we have
\be \lb{rn21}
\psi (x)=\frac{g}{\left( 1+\frac{R_{H}^{n-2}}{r_{0}^{n-2}}\right) \Gamma^{n/2-1}(x,x_0)} .
\ee
It is obvious that expression (\ref{rn21}) is regular at the horizon $r=0$ and at the infinity.

\section{Conclusion}

In this work, we have characterized the static, spherically symmetric metrics
in which the elementary solution of the electrostatic equation or of the static scalar equation
can be expressed in closed form (\ref{02}). When this metric represents a 
black hole with a non-degenerate horizon, we have found in the electrostatic case a
radical situation : it is only possible in four dimensions. This general study
will avoid the researches for each possible black hole. 

The situation is different for a static scalar field in non-degenerated black holes.
There exists metrics in $n+1$  dimensions having non-degenerated horizon 
in which the elementary solution for scalar field has form (\ref{02}). 
These are not the black holes of the Einstein theory in $n+1$ dimensions,
in particular the Ricci scalar curvature does not vanish. We are confined ourselves to
minimally coupled scalar field. We have proved in $n+1$ dimensions with $n$ odd that the
scalar field generated by a scalar charge, held fixed in these black holes, coincides
with the elementary solution up to a multiplicative factor. This suggests very strongly 
that there is no scalar self-energy in these cases \cite{mayo,wiseman,zelnikov}.

We have also determined the electrostatic potential generated by an electric charge in
the extremal Reissner-Norstr\"{o}m black hole in $n+1$ dimensions. Of course, we must 
in principle consider
a perturbation in the framework of the Einstein-Maxwell theory in order to be coherent, like
in the case $n=3$ \cite{bicak}. 
However, we can envisage a massless vector field different from the electromagnetic field of the
background charged metric.
We have found that the electrostatic potential does not coincide with the elementary
solution. The homogeneous solution that we must add to eliminate the electric flux
through the horizon induces an electrostatic self-energy.

Unfortunately, we are not in a position to apply 
the idea of Bekenstein and Mayo \cite{bekenstein} and Hod \cite{hod} on the entropy bound 
for a charged object in $n+1$ dimensions because the surface of gravity of
the extremal black holes vanishes. 
Hence we cannot test on our exact solution the suggestion of  
Cai et al \cite{cai} on the entropy bound for a charged object in higher dimensions.

We have furthermore given in the case I for $n=3$, the general solution for $\xi$, 
independently of the existence of an horizon, which yields form (\ref{02})
of the elementary solution. The differential equations in the other cases are easy to solve. 

\appendix
\section{Case I: general solution for $n=3$}

In the case $n=3$ equation (\ref{32}) reduces to
\be \lb{32n3}
\xi '' -\frac{2a^2}{r(r^2-a^2)}\xi ' +\frac{Aa^2}{(r^2-a^2)^2}\xi =0
\ee
We have found the general solution to equation (\ref{32n3}). The expression depends
on the value of the constant $A$. For $A<4$, we have 
\bea \lb{511g}
& & \xi_1(r)=\frac{r^2-a^2}{r}\left( \frac{r+a}{r-a}\right)^{-\sqrt{1-A/4}} , \nonumber \\
& & \xi_2(r)=\frac{r^2-a^2}{r}\left( \frac{r+a}{r-a}\right)^{\sqrt{1-A/4}} ,
\eea
which corrsponds at $r=a$ respectively to the indice $k$, given by $k=1+\sqrt{1-A/4}$, and 
to the indice $2-k$.
For $A=4$, we have a solution with the indice $1$ at $r=a$ and a solution with a
logarithmic term
\be \lb{5a0}
\xi_1(r)=\frac{r^2-a^2}{r} ,\quad
\xi_2(r)=\frac{r^2-a^2}{r}\ln \left( \frac{r-a}{r+a}\right) .
\ee
For $A>4$, the indice $k$ at $r=a$ is not real and therefore we take real solutions under the form
\bea \lb{511d}
& & \zeta_1(r)=\frac{r^2-a^2}{r}\cos \left[ \sqrt{A/4-1}\ln \left( \frac{r-a}{r+a}\right) \right] ,\nonumber \\
& & \zeta_2(r)=\frac{r^2-a^2}{r}\sin \left[ \sqrt{A/4-1}\ln \left( \frac{r-a}{r+a}\right) \right] .
\eea

In order to be complete, we give the expression of $F$ in this case.
For $n=3$, equation (\ref{ab9}) becomes
\be \lb{ab9n3}
4(s^2+s) F''+(12s+6) F' +AF =0 .
\ee
The solution to (\ref{ab9n3}) must admit a development in the Hadamard sense for $s$. 
For $A\leq 4$, we obtain
\bea \lb{27} 
& & F(s)=\frac{1}{2\sqrt{s(s+1)}} \times \nonumber \\
& & \left[ \left( 1+2s+2\sqrt{s(s+1)}\right)^{\sqrt{1-A/4}}+\left( 1+2s+2\sqrt{s(s+1)}\right)^{-\sqrt{1-A/4}} \right] .
\eea
For $A>4$, we obtain
\be \lb{27b}
F(s)=\frac{1}{\sqrt{s(s+1)}}\cos \left[ \sqrt{A/4-1}\ln \left( 1+2s+2\sqrt{s(s+1)}\right) \right] .
\ee
We have a development in the Hadamard sense since (\ref{27}) et (\ref{27b}) have the expansion
\be \lb{exp}
F(s)\sim \frac{1}{\sqrt{s}}\left( 1+\frac{3-A}{2}s+\ldots \right) \quad {\rm quand} \; s\rightarrow 0 .
\ee
In a previous work in the Brans-Dicke theory \cite{linet3}, we have given a solution to (\ref{ab9n3})
which was not a solution in the Hadamard sense.


\end{document}